\documentclass[superscriptaddress, nofootinbib, twocolumn, amsmath,amssymb, aps, pra, notitlepage, longbibliography]{revtex4-1}

\usepackage{dcolumn}
\usepackage{bm}
\usepackage{epsfig}
\usepackage{graphicx}
\usepackage{latexsym}
\usepackage{amsfonts}
\usepackage{setspace}
\usepackage{graphicx}
\usepackage{amsmath}
\usepackage{verbatim}
\usepackage{color}
\usepackage{hyperref}
\usepackage{nccmath}

\usepackage{siunitx}

\begin{document}

\title{Highly localized Brillouin scattering response in a photonic integrated circuit}

\author{Atiyeh Zarifi$^{1,2}$, Birgit Stiller$^{1,2,\ast}$, Moritz Merklein$^{1,2}$, Neuton Li$^{1}$, Khu Vu$^{3}$,  Duk-Yong Choi$^{3}$, Pan Ma$^{3}$, Stephen J. Madden$^{3}$ and Benjamin J. Eggleton$^{1,2}$\\
\small{ \textcolor{white}{blanc\\}
$^{1}$Centre for Ultrahigh bandwidth Devices for Optical Systems (CUDOS), Institute of Photonics and Optical Science (IPOS), School of Physics, University of Sydney, Sydney, New South Wales 2006, Australia.\\
$^{2}$Australian Institute for Nanoscale Science and Technology (AINST), University of Sydney, Sydney NSW 2006, Australia.\\
$^{3}$Centre for Ultrahigh bandwidth Devices for Optical Systems (CUDOS), Laser Physics Centre, Research School of Physics and Engineering, Australian National University, Canberra, Australian Capital Terrritory 0200, Australia.\\
$^{\ast}$birgit.stiller@sydney.edu.au\\}}

\begin{abstract} 

The interaction of optical and acoustic waves via stimulated Brillouin scattering (SBS) has recently reached on-chip platforms, which has opened new fields of applications ranging from integrated microwave photonics and on-chip narrow-linewidth lasers, to phonon-based optical delay and signal processing schemes. Since SBS is an effect that scales exponentially with interaction length, on-chip implementation on a short length scale is challenging, requiring carefully designed waveguides with optimized opto-acoustic overlap. In this work, we use the principle of Brillouin optical correlation domain analysis (BOCDA) to locally measure  the SBS spectrum with high spatial resolution of 800 \textbf{\SI{}{\micro\meter}} and perform a distributed measurement of the Brillouin spectrum along a spiral waveguide in a photonic integrated circuit (PIC). This approach gives access to local opto-acoustic properties of the waveguides, including the Brillouin frequency shift (BFS) and linewidth, essential information for the further development of high quality photonic-phononic waveguides for SBS applications.

\end{abstract}

\maketitle

\section{Introduction}

Stimulated Brillouin scattering (SBS) is an opto-acoustic interaction, in which the energy of the pump wave transfers into a frequency down-shifted probe wave (Stokes) through an acoustic wave. SBS enables powerful applications such as narrow linewidth lasers \cite{Loh2015,Lee2012,Kabakova2013,Stiller2013d}, optical delay lines \cite{Thevenaz2008,Zhu2007,Merklein2016f}, temperature and strain sensors \cite{Song2007,Galindez-Jamioy2012}, microwave generation \cite{Li2013,Merklein2016} and signal processing \cite{Kobyakov2010b,Marpaung2013,Choudhary2017}. In particular, the possibility to generate SBS on-chip opens a new paradigm for compact integrated devices in a small-footprint \cite{Choudhary2017,Merklein2016,Merklein2016g}. Harnessing SBS on-chip, however, is challenging since SBS scales exponentially with the interaction length \cite{agrawal2007nonlinear}. To achieve strong opto-acoustic coupling, it is necessary to confine the optical and acoustic mode simultaneously in a small cross-section \cite{Poulton2013}. One possibility is to use soft-glass waveguides sandwiched between a more rigid cladding material \cite{Pant2011}, however in recent years more complex structures were put forward to generate strong on-chip opto-acoustic overlap, such as under etched silicon waveguides \cite{VanLaer2015,Kittlaus2016}, hybrid silicon-siliconnitride membranes \cite{Shin2013,Shin2015}, fully suspended nanowires \cite{VanLaer2015}, bandgap engineered soft-glass waveguides \cite{Merklein2015} or multi-material hybrid circuits \cite{Casas-Bedoya2016,Morrison:17}. The high complexity of the waveguides that guide optical and acoustic waves, and the sensitivity of these modes to even small variations in geometry \cite{wolff2016brillouin,Hotate2012c,Stiller2012} requires a technique to probe the opto-acoustic coupling strength on a sub-millimeter length scale. These new insights will help to better understand local parameters effecting the overall SBS gain response, ensuring high-performance and high yield of the photonic waveguides.

Numerous distributed SBS measurement techniques have been developed, including Brillouin optical time domain analysis (BOTDA) \cite{Thevenaz2010,Bao1993,Nikles1996}, Brillouin echo distributed sensing (BEDS) \cite{Foaleng:10,Beugnot2011} and Brillouin optical correlation domain analysis (BOCDA) \cite{Hotate2000,Hotate2013,Motil2016,Hasegawa1999,Zadok2012,Antman2013,Cohen2014a,Cohen2014}. BOTDA has limited spatial resolution, since the pump pulses should not be shorter than the phonon lifetime (corresponding to 1 m spatial resolution in optical fiber). BEDS proposal was put forward to overcome this limitation relying on a short $\pi$ phase shift applied to the continuous wave (CW) pump instead of a pulsed pump, which improves the spatial resolution down to \SI{1}{\cm} \cite{Thevenaz2010,Foaleng:10,Beugnot2011}. 
BOCDA, inspired by the field of radar, enables millimeter spatial resolution SBS measurement of the waveguide. Similar to radar, which deals with detecting the position and the velocity of a moving object \cite{Levanon2004} relative to a source, BOCDA enables localized detection of the Stokes wave, whose frequency is Doppler shifted relative to the pump wave. The first BOCDA technique \cite{Hotate2000,Hasegawa1999} was based on the broad-spectrum frequency-modulated pump and probe waves. Following this initial demonstration, different variations of BOCDA have been introduced, including random bit phase-modulated pump and probe \cite{Zadok2012}, Golomb-code modulated pump and probe \cite{Antman2013} and noise-based correlation technique \cite{Cohen2014a,Cohen2014}. All these techniques rely on the fact that the Fourier transform of the pump and the probe spectrum product is a spatially localized correlation peak, whose linewidth defines the spatial resolution of the system. The realization of the broad-spectrum pump and probe through the frequency and phase modulation adds to the complexity of the experiment, whereas the noise-based correlation technique relies only on the amplified spontaneous emission (ASE) of an Erbium doped fiber \cite{Cohen2014a}, which offers simplicity and high spatial resolution but suffers from limited signal to noise ratio (SNR).

In this work, we achieve a record on-chip spatial resolution using a noise-based BOCDA measurement with a lock-in amplifier (LIA) to improve the SNR. We spatially resolve the Brillouin response of a chalcogenide photonic waveguide on a chip with a high spatial resolution of \SI{800}{\micro\meter}, which we believe is the smallest section over which SBS has been observed in a planar waveguide. This setup employs the ASE of an Erbium doped fiber to resolve features such as waveguide thickness and the resultant effective refractive index change of an $\textup{As}_{2}\textup{S}_{3}$ photonic waveguide. This work provides the basis for the understanding of a very local interaction of optical and acoustic waves and the design of novel waveguide structures.

\section{principle of operation}

SBS is an inelastic scattering effect, in which two counter-propagating optical pump and probe waves generate a moving index grating in the medium, which travels with the acoustic velocity ($\textnormal{v} _{a}$). The pump wave is backscattered by the moving index grating, which results in a frequency down-shifted Stokes wave in the backward direction. The frequency difference between the pump and the Stokes is called the Brillouin frequency shift (BFS) $\Omega_{B}$ and is defined as \cite{agrawal2007nonlinear,Robert.WBoyd2007}: 

\begin{equation}
\Omega _{B}=\frac{2n_\textrm{eff}\textnormal{v}_{a}}{\lambda }, 
\label{eq:delta nu}
\end{equation}
where $n_\textup{eff}$ is the effective refractive index of the medium and $\lambda$ is the pump wavelength.
The BFS depends upon the optical material as well as the waveguide dimensions and environmental conditions such as temperature and strain, which makes SBS suitable for sensing purposes. In order to employ SBS as a distributed sensing approach, a localize BFS measurement is required. Among the different distributed SBS techniques introduced in the previous section, we select BOCDA since it provides the highest spatial resolution down to mm-scale,which is the preferred method for mapping cm-scale integrated photonic circuits. 

\begin{figure}[t]
\centering
\includegraphics[width=\linewidth]{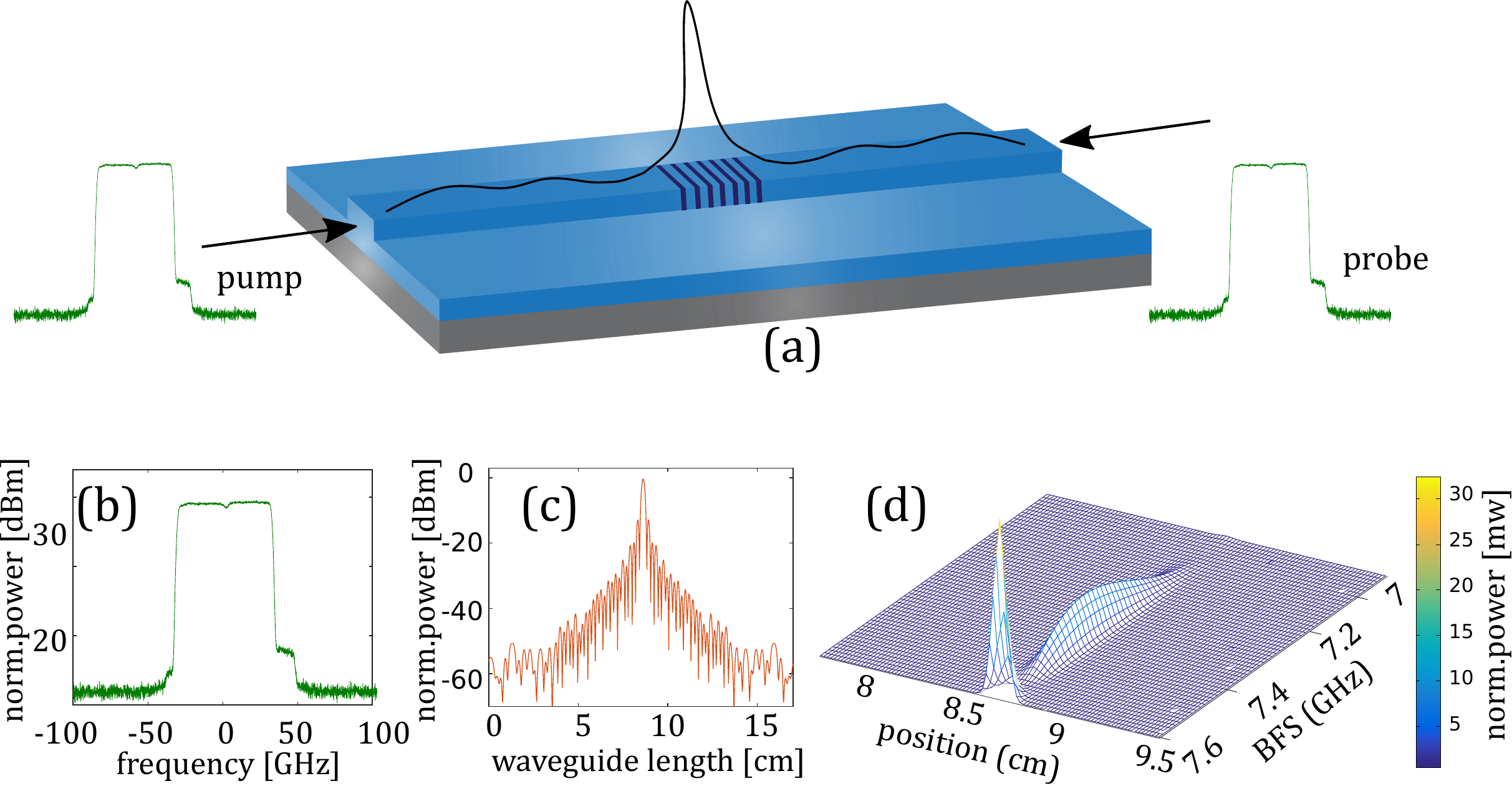}
\caption{a) Correlation peak in the waveguide as a result of broad-spectrum pump and probe. b) 80 GHz ASE spectrum and c) the  autocorrelation function of its complex envelope u(t) in dB scale along the waveguide. d) Ambiguity function of an 80 GHz square ASE spectrum in a $\textup{As}_{2}\textup{S}_{3}$ waveguide, the colormap is scaled with the normalized linear power of the ambiguity function.}
\label{fig:correlation peak}
\end{figure}

The key to realize a localized SBS measurement using BOCDA is that unlike the conventional SBS measurement technique, which relies on the coherent pump and probe signals, BOCDA employs a highly non-coherent signal as the pump and the probe. In general, the cross-correlation function between the pump and the probe signals (also called as the auto-correlation function, since the pump and the probe signals are driven from the same source) defines the level of the localization of the SBS response. For an integrated SBS measurement, the coherent pump and probe signals create a constant cross-correlation function along the medium, therefore the SBS response is the accumulative gain collected from the entire length of the medium. In BOCDA, however, the cross-correlation between the non-coherent pump and probe signals gives rise to a spatially localized correlation peak, which confines the SBS response into a narrow region and suppresses it elsewhere in the medium. This gives access to a localized SBS response and is illustrated in Fig. \ref{fig:correlation peak}(a). 

A filtered ASE source with a bandwidth of 80 GHz as the non-coherent source and the auto-correlation function of its complex envelope u(t) are shown in Fig. \ref{fig:correlation peak}(b) and Fig. \ref{fig:correlation peak}(c), respectively. The bandwidth $\Delta \textnormal{f}$ of the ASE spectrum defines its coherence time $\Delta t$ by:  $\Delta t\Delta \textnormal{f} \approx 1$. Therefore, the spatial resolution of the localized response can be adjusted by changing the ASE bandwidth.

The overall SBS gain g is obtained by adding together the individual local Brillouin gain spectrum $g_{B}$ spatially weighted by the cross-correlation function over the entire length of the waveguide using the following equation \cite{Hotate2000}: 

\begin{equation}
g=\frac{v_{g}\bar{P}_{1}}{A_\textup{eff}T}\int_{-\infty }^{\infty}d\tau\int_{-\infty }^{\infty} g_{B}(\tau,\nu )\left | \chi(\tau,\nu)  \right |^{2} d\nu,
\label{eq:overal gain}
\end{equation}
where $v_{g}$ is the group velocity of light in the medium, $\bar{P}_{1}$ is the average pump power, $A_\textup{eff}$ is the waveguide effective area, T is the period of the optical wave and $\chi(\tau,\nu)$ is a two dimensional function, which defines the cross-correlation between the complex envelope of the pump $u(t)$ and the probe $u(t-\tau)$ as a function of the delay ($\tau$) and the frequency shift ($\nu$) between the pump and the probe. $\chi (\tau ,\nu)$ is calculated using the following equation \cite{Levanon2004,Hotate2000}: 

\begin{equation}
\chi (\tau ,\nu)= \frac{nc\epsilon_{0}A_\textup{eff}}{2\sqrt{\bar{P_1}\bar{P_2}}} \int_{-\infty }^{\infty }u(t)u^{\ast }(t-\tau)\textup{exp}(i2\pi\nu t)dt,
\label{eq:AF}
\end{equation}
where n is the refractive index of the medium, c is the speed of light in the vacuum, $\epsilon_{0}$ is the free space permittivity and $\bar{P}_{2}$ is the average probe power. Equation \ref{eq:AF} is also known as the ambiguity function \cite{Levanon2004}, since it introduces a degree of ambiguity in the measured Brillouin gain spectrum. A simulation of the ambiguity function for the \SI{80}{\giga\hertz} ASE spectrum pump and probe is shown in Fig. \ref{fig:correlation peak}(d).

In this simulation, the medium is a \SI{17.5}{\cm} chalcogenide waveguide with the BFS of \SI{7.6}{\giga\hertz}, and the pump and the probe arm are assumed to have the same length (zero delay between the pump and the probe). The delay between the pump and the probe can be translated into the position in the waveguide using the relation: $x = \frac{1}{2}(v_{g}\tau +L)$, where $v_{g}$ is the group velocity in the waveguide and L is the total length of the waveguide. As it can be seen in  Fig. \ref{fig:correlation peak}(d), the correlation peak occurs in the middle of the waveguide (plotted in linear scale), and the frequency response is maximum at \SI{7.6}{\giga\hertz}. 

In our experiments, a filtered ASE of an Erbium doped fiber is used as the common source for the pump and the probe. The correlation peak position is moved along the medium using a delay line to change the relative delay between the pump and the probe arms. The integral sum of the local Brillouin gain spectrum multiplied by the local spectrum of the ambiguity function over the length of the medium is measured for each delay step, which is referred to as the local SBS response in this document. This product is maximum at the correlation peak and is suppressed by the shape of the correlation function at any other point. The contribution of these points add to the ambiguity (noise) of the overall Brillouin gain spectrum.

\section{Experiment and discussion}
\subsection{Experimental Setup}

\begin{figure}[t]
\centering
\includegraphics[width=\linewidth]{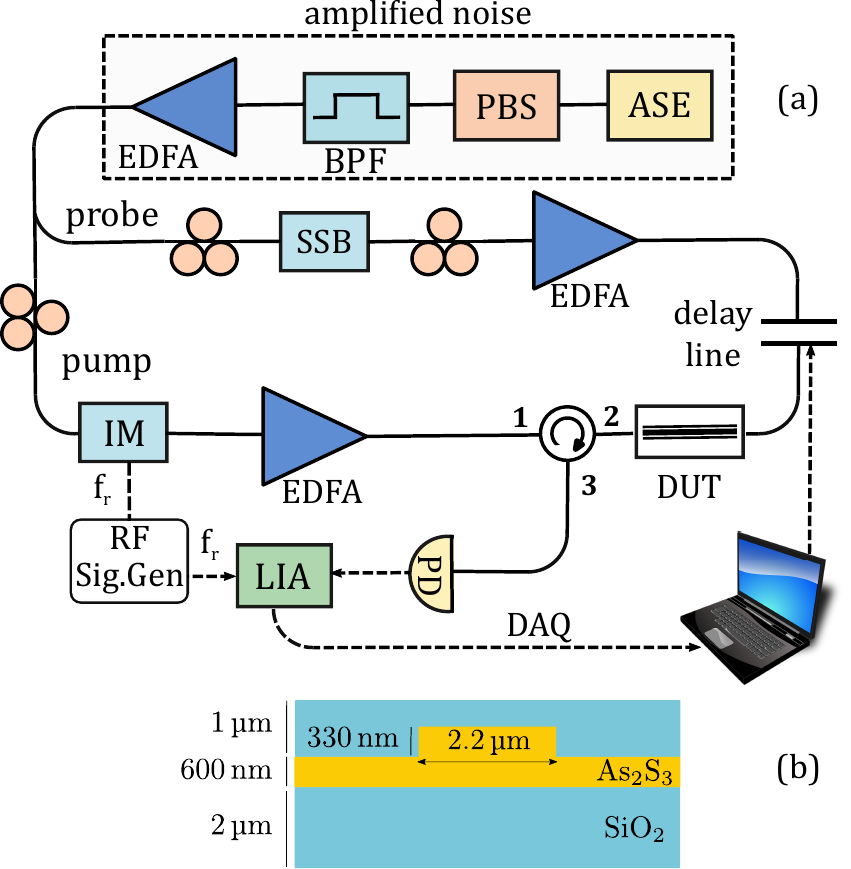}
\caption{a) Setup for BOCDA measurement. LIA: lock-in amplifier, BPF: band-pass filter,
PBS: polarization beam splitter,
IM: intensity modulator,
SSB: single side band modulator,
PD: photo-detector,
DUT: device under test,
$\textup{f}_{r}$: reference frequency,
RF Sig. Gen.: radio frequency signal generator,
DAQ: data acquisition.
b) Cross section of the $\textup{As}_{2}\textup{S}_{3}$ waveguide.}
\label{fig:updated_setup}
\end{figure}

The experimental setup is shown in Fig. \ref{fig:updated_setup}(a). The ASE spectrum of an Erbium doped fiber passes through a polarization beam splitter (PBS) and is filtered to a square-shape spectrum. The bandwidth of the band-pass filter varies from \SI{25}{\giga\hertz} to \SI{89}{\giga\hertz} depending on the desired spatial resolution. An Erbium doped fiber amplifier (EDFA) is used to pre-amplify the polarized filtered ASE, which is then divided between the pump and the probe arms with a power ratio of \SI{30}{\percent} to \SI{70}{\percent}, respectively. 
 
The pump signal is modulated by a Mach-Zehnder modulator (MZM) using square pulses of \SI{500}{\nano\second} width and \SI{5}{\percent} duty cycle. The same radio-frequency (RF) source that drives the MZM, triggers the lock-in amplifier (LIA) with the reference frequency of \SI{100}{\kilo\hertz}. In the probe arm, a dual-parallel Mach-Zehnder modulator (DPMZM) is used to generate a single sideband (SSB), down-shifted by the BFS (\SI{10.8}{\giga\hertz} for SMF and approximately \SI{7.6}{\giga\hertz} for chalcogenide waveguides). This technique results in \SI{25}{\decibel} of side-band and carrier suppression, which is critical to improve the SNR.

The SSB is swept over a \SI{350}{\mega\hertz} frequency span around the BFS using an automated RF signal generator to measure the SBS gain spectrum.
The delay line is swept automatically to change the delay between the pump and the probe arms and therefore, changing the position of the correlation peak in the medium. The pump and the probe signals counter-propagate through the medium and the backscattered signal (Stokes) is collected at port 3 of the circulator. At the LIA, only frequency components which match the frequency of the pump pulses (reference frequency) are detected and amplified and the rest of the ASE spectrum coming from the probe is rejected; this will improve the SNR of the measurement.

\subsection{Signal to noise ratio (SNR)}

In this setup we directly measure the probe amplification using the LIA. The measured backscattered signal has a component in the reference frequency (\SI{100}{\kilo\hertz}), which is a combination of the pump back reflection and the amplified probe. The contribution of the pump back reflection is always present in the measurement, which in turn increases the noise level. This pump back reflection cannot be filtered because the pump and the amplified probe have 75\%  spectral overlap and by filtering the pump, a major part of the SBS response is also removed. A common approach to remove the pump back reflection is to chop both the pump and the probe at two different frequencies and to use the difference frequency as the LIA reference frequency \cite{Kwang-YongSong2006}. Assuming the pump back reflection contribution is equal for all the points in a single measurement, we define the SNR as the difference between the peak intensity and the noise floor divided by the noise floor for a single local measurement. In our experiment, the SNR varies from 0.03 to 0.18 depending on the measurement.


\subsection{Scanning The Waveguide}

We aim to fully detect and resolve a short $\textup{As}_{2}\textup{S}_{3}$ rib waveguide with \SI{2.2}{\micro\meter} width and \SI{930}{\nano\meter} thickness as shown in Fig. \ref{fig:updated_setup}(b). 
The light is coupled into and out of the waveguide using lensed fibers with approximate coupling loss of 4.2 $\pm$ 0.2 dB per facet. The propagation loss is about \SI{0.2}{\decibel\per\cm} and the total loss of the waveguide is 9 $\pm$ 0.5 dB at the pump peak power of 27 dBm.
The filter bandwidth is set to \SI{25}{\giga\hertz} corresponding to \SI{2.5}{\mm} spatial resolution in the waveguide according to:

\begin{equation}
\Delta x \approx \frac{1}{2}v_{g}\Delta t, 
\label{eq:spatial res}
\end{equation}
where $v_{g}=\frac{c}{n_\textup{g}}$ and $n_\textup{g}$ is the group index of the waveguide.
 
The delay line and the RF signal generator are controlled by a computer program; the delay line takes \SI{1}{\mm} steps at every \SI{5}{\second} and the RF signal generator sweeps the probe signal over \SI{350}{\mega\hertz} span with \SI{2,5}{\mega\hertz} spectral resolution for each delay step. For every frequency, the backscattered signal is collected at port 3 of the circulator and is measured by the LIA. 
\begin{figure}[t]
\centering
\includegraphics[width=\linewidth]{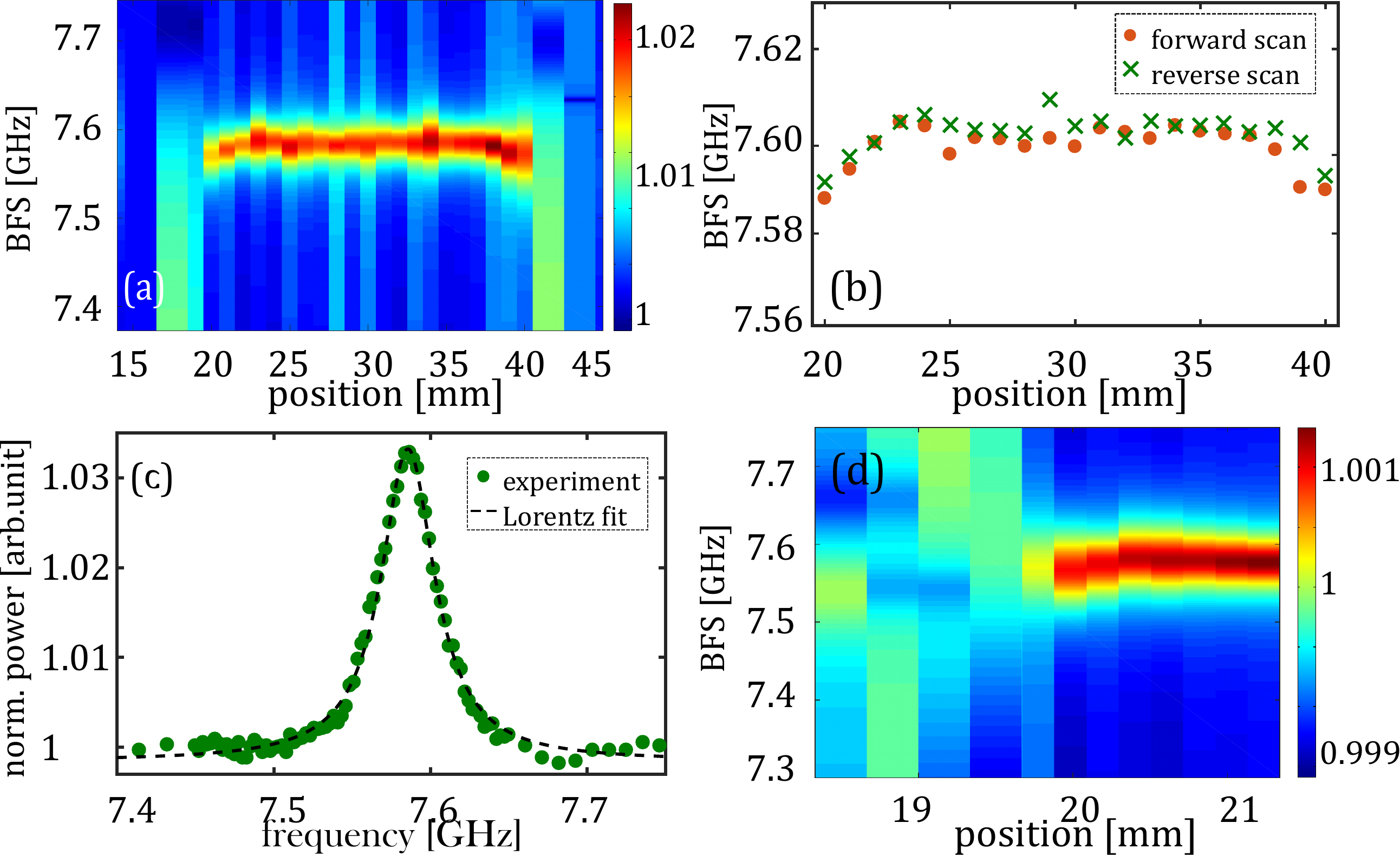}
\caption{a) Normalized, Lorentz-fitted BOCDA measurement of a short $\textup{As}_{2}\textup{S}_{3}$ waveguide (the colormap plot shows the linear normalized power). b) BFS of the forward and backward scan. c) A local Brillouin spectrum at \SI{1.4}{\cm} from the waveguide facet with a Lorentzian fit. d) Scan of the waveguide facet with \SI{800}{\um} spatial resolution, averaged over four traces (the colormap shows the linear normalized power).}
\label{fig:short wg}
\end{figure}
By moving the correlation peak through the waveguide using the delay line, a map of the local Brillouin response is created as shown in Fig. \ref{fig:short wg}(a). The region over which the local Brillouin responses are detected is approximately \SI{20.8}{\mm}, corresponding to the length of the waveguide. 
By fitting the local responses with a Lorentzian profile, a map of BFS over the length of the waveguide is established. The BFS over this region is fairly consistent with a mean value of \SI{7.58}{\giga\hertz} and standard deviation of \SI{3.2}{\mega\hertz}, which confirms the uniformity of the waveguide. In order to confirm this measurement, the waveguide is scanned in a reverse direction; that is the pump and the probe directions are swapped. This measurement also shows consistent BFS over the waveguide length. The BFS associated with the two measurements are shown in Fig. \ref{fig:short wg}(b). 

A local Brillouin response at a random position in the waveguide (\SI{1.4}{\cm} from the waveguide input facet) is shown in Fig. \ref{fig:short wg}(c). A Lorentzian profile is fitted to the measured points, which has a linewidth of \SI{41}{\mega\hertz} and a maximum at \SI{7.58}{\giga\hertz}, corresponding to the linewidth and the BFS of the local SBS gain spectrum. The measured local gain at this position is approximately \SI{0.15}{\decibel}, and the SNR is 0.03.

The filter bandwidth is then increased to \SI{80}{\giga\hertz}, which corresponds to a spatial resolution of \SI{800}{\um} in the waveguide. The edge of the waveguide is resolved using this high spatial resolution setting as it is shown in Fig. \ref{fig:short wg}(d). Since the SBS interaction length is only \SI{800}{\um}, an averaging over four measurements is required to resolve this region. The delay steps in the delay line are set to 1 mm in free space and the scanning range is over 3 mm length of the waveguide. As the correlation peak travels from the lensed fiber into the chalcogenide waveguide, the optical field experiences different effective refractive indices. This means that for a fixed delay step in free space (\SI{1}{\mm}), the delay steps outside the waveguide are longer (\SI{0.33}{\mm}) than the delay steps inside the waveguide (\SI{0.2}{\mm}). This feature can be seen in Fig. \ref{fig:short wg}(d).

\subsection{Waveguide Characterization}

The second experiment aims to look at the uniformity of a long spiral waveguide (\SI{17.5}{\cm} long), which consists of several \ang{180} bends (with bend radius of approximately \SI{200}{\um}) as well as straight regions. The waveguide is in the form of a rib waveguide and is \SI{2.4}{\micro\meter} wide and \SI{930}{\nano\meter} thick.

\begin{figure}[t]
\centering
\includegraphics[width=\linewidth]{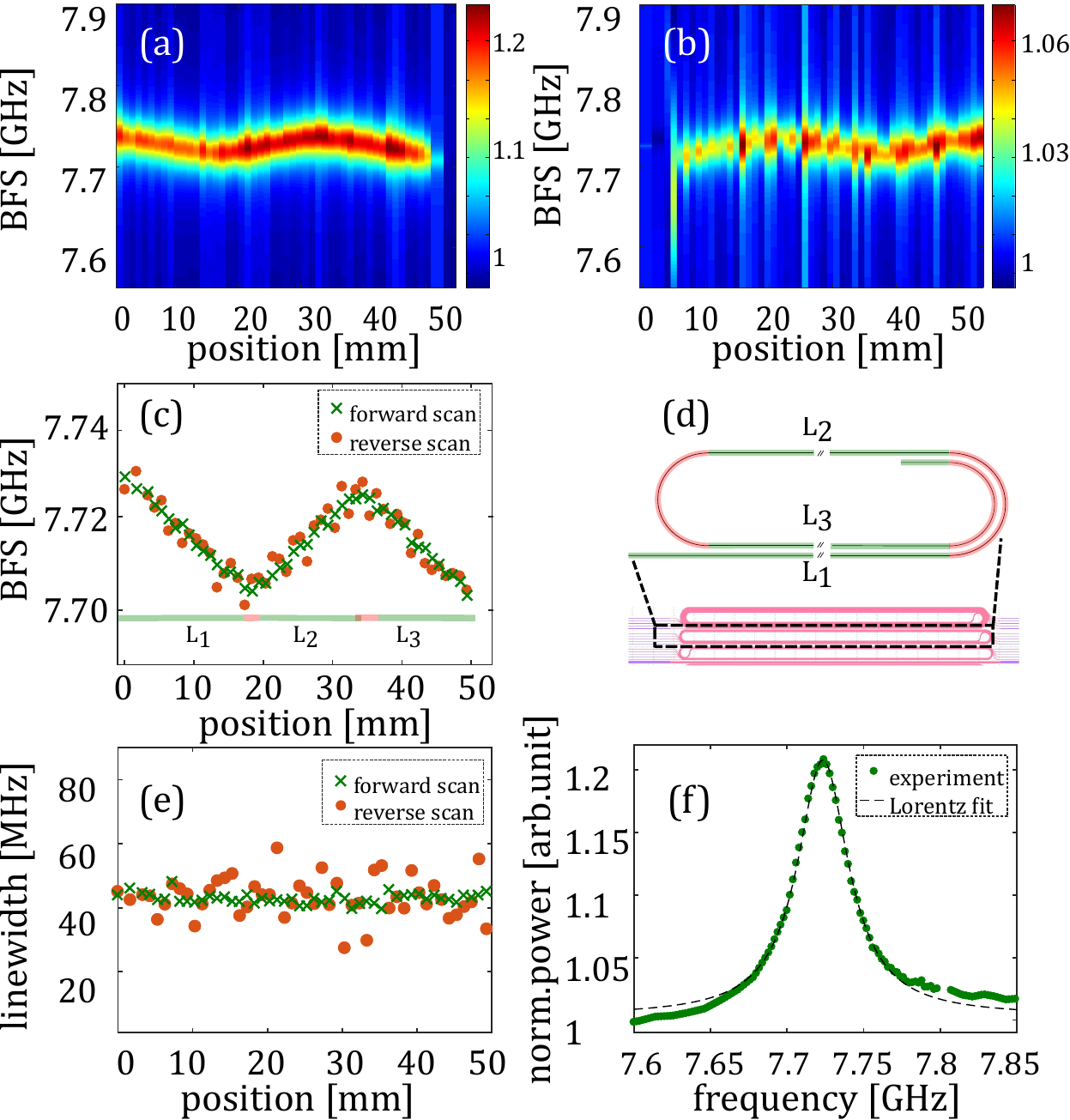}
\caption{BOCDA scan of the long waveguide in a) forward direction
and b) reverse direction (the colormap shows the linear normalized power). c) BFS of the forward and the
backward scan. d) the mask layout and the schematic of the scanned region in the waveguide. e) Local SBS linewidth of the forward and reverse scan. f) Local SBS response at position \SI{2}{\cm}  from the front facet in the forward direction.}
\label{fig:long}
\end{figure}

The filter bandwidth is set to \SI{62.5}{\giga\hertz} corresponding to \SI{1}{\mm} spatial resolution in the waveguide. The waveguide insertion loss is 15 $\pm$ 1 dB and the pump peak power before coupling is set to \SI{29}{\textnormal{dBm}}. The spectral resolution of the measurement is \SI{2.5}{\mega\hertz} and each local measurement takes \SI{5}{\second} to complete. Fig. \ref{fig:long}(a) shows a map of local Brillouin responses over the first \SI{5}{\cm} of the waveguide, which is closer to the pump arm. As it is plotted in Fig. \ref{fig:long}(a), the local BFS gradually changes in an oscillating pattern as we scan through the waveguide. 
In order to confirm this measurement, the waveguide is scanned in the reverse direction by swapping the pump and the probe connections to the waveguide. As it is shown in Fig. \ref{fig:long}(b), which is a mirrored image of the first scan, the same variations in the local BFS is observed when the direction of the scan is changed. From Fig. \ref{fig:long}(b), it can be seen that the SNR of the local responses is poor compared to the first scan. This can be explained by taking into account the effect of the propagation loss that the pump experiences, which is higher in the second measurement compared to the first one, since the pump has to propagate a longer distance to reach the same point as in the first measurement. 

Fig. \ref{fig:long}(c) confirms that the local BFS in the forward and the reverse scan match. In addition, a local BFS variation of \SI{22}{\mega\hertz} is observed in this plot. This suggests that the optical mode experiences a varying effective refractive index as it propagates through the waveguide, according to Equation \ref{eq:delta nu}.

The most likely explanation for the BFS variation is the fabrication imperfections caused by the non-uniformity in the deposition of the $\textup{As}_{2}\textup{S}_{3}$ layer, which results in a non-uniform waveguide thickness. 
A commercial-grade simulator eigenmode solver and propagator was used to calculate the effect of waveguide thickness on the BFS \cite{Lumerical:2009:Misc}. The simulation result indicates that the waveguide thickness variation of 5\% corresponding to the effective refractive index change of \SI{0.002}, will cause a BFS change of \SI{25}{\mega\hertz}. This variation is close to the value we observed in the experiment (\SI{22}{\mega\hertz}) and is consistent with the expected fabrication uniformity for the $\textup{As}_{2}\textup{S}_{3}$ layer; since the waveguide is located near the edge of the wafer, where the deposition variations are more pronounced.

By matching the BFS to the waveguide layout as shown in Fig. \ref{fig:long}(d), it can be seen that as the correlation peak is moved from the edge of the waveguide toward the bends and travels back to the edge, the BFS changes accordingly. This confirms that the waveguide is slightly thinner at the edge of the waveguide and becomes thicker at the center.

The effect of the bends on the effective refractive index change and consequently the BFS change is also studied; the numerical simulation indicates \SI{3}{\mega\hertz} BFS change due to the bends for the fundamental TE mode (which is the dominant mode in the waveguide) and \SI{120}{\mega\hertz} change in the BFS as a result of coupling to the higher order mode. Comparing these values with the experimental observation, confirms that the BFS variation is not due to the bends and that the optical mode remains unaffected by the bends. 

The linewidth of the local SBS responses along the waveguide is shown in Fig. \ref{fig:long}(e). The standard deviation is \SI{1.53}{\mega\hertz} and \SI{6.03}{\mega\hertz} for the forward and the backward scan, respectively. The local Brillouin response and its Lorentzian fit at position \SI{2}{\cm} from the edge of the waveguide for the forward measurement are depicted in Fig. \ref{fig:long}(f). The local response has a linewidth of \SI{42.7}{\mega\hertz} and a BFS of \SI{7.72}{\giga\hertz} as shown in Fig. \ref{fig:long}(f). The local gain at this position is measured to be \SI{0.9}{\decibel}, and the SNR is 0.18.

\subsection{Spatial Resolution Confirmation}

In order to confirm the spatial resolution of the BOCDA setup, detection of a \SI{1}{\mm} long dispersion shifted fiber (DSF) spliced in between the two pieces of single mode fiber (SMF) with slightly different BFS is demonstrated. 

\begin{figure}[t]
\centering
\includegraphics[width=\linewidth]{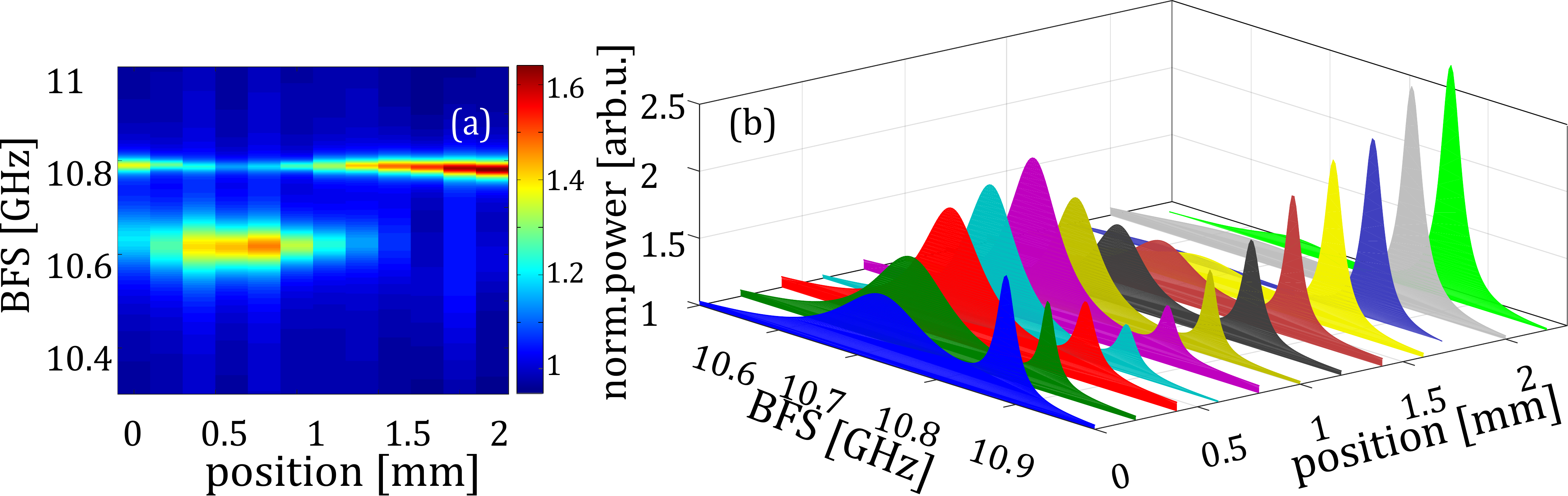}
\caption{a) Scan of a \SI{1}{\mm} piece of DSF spliced in between the SMF fiber with \SI{1.1}{\mm} spatial resolution (the colormap shows the linear normalized power). b) Local SBS response at different positions of the sample.}
\label{fig:DSF}
\end{figure}

The ASE bandwidth is set to \SI{89}{\giga\hertz}, which corresponds to \SI{1.1}{\mm} spatial resolution in silica fiber. The correlation peak is moved by \SI{0.2}{\mm} steps in fiber. As it moves from the SMF into the DSF and back to the SMF again, the BFS of the local Brillouin responses changes from \SI{10.89}{\giga\hertz} to \SI{10.68}{\giga\hertz} and back to \SI{10.89}{\giga\hertz} as shown in Fig. \ref{fig:DSF}(a). Fig. \ref{fig:DSF}(b) shows the appearance and disappearance of the DSF peak over the short region of 1 mm. As it is observed in Fig. \ref{fig:DSF}(b), the SMF peak is present in all the traces since part of the correlation peak always overlap with the SMF. However, the amplitude of the SMF peak is the lowest when the correlation peak is entirely inside the DSF, and has minimum overlap with the SMF. The SNR in this measurement is low because silica fiber has lower SBS gain compared to the $\textup{As}_{2}\textup{S}_{3}$ waveguide. Therefore, we averaged the measurement over four traces to fully recover the DSF piece.

\section{conclusion}

A distributed SBS measurement setup based on the principle of BOCDA with record spatial resolution of \SI{800}{\um} is presented. Our system gives access to the degree of uniformity of $\textup{As}_{2}\textup{S}_{3}$ waveguide, which was confirmed in a straight and a spiral waveguide. As a proof of principle demonstration, we resolved 5\% variations in waveguide thickness, which is in agreement with expected fabrication tolerances. We also confirmed the spatial resolution of the system by detecting a 1 mm piece of DSF which was spliced in between two pieces of SMF with slightly different BFS. This demonstration presents the first on-chip SBS response measurement with sub-mm resolution and opens up new opportunities to discover local fundamental opto-acoustic effects in SBS integrated platforms, which have not yet been studied closely.

\section*{Funding Information}
This work was funded by the Australian Research Council (ARC) Laureate Fellowship (FL120100029) and the Centre of Excellence program (CUDOS
CE110001010).

\section*{Acknowledgments}

We acknowledge the support of the Australian National Fabrication Facility (ANFF) ACT Node in carrying out this research.
The authors would like to thank Mr. Blair Morrison for his insightful discussions during this work.
   
\bibliography{optica_ref.bib}


\end{document}